\documentclass[a4paper,11pt,sort&compress]{elsarticle}
\usepackage[utf8]{inputenc}
\usepackage[margin=1in]{geometry}
\usepackage{amsmath}
\usepackage{amssymb}
\usepackage{graphicx}
\usepackage{xcolor}
\usepackage[caption=false]{subfig}
\usepackage{hyperref}

\newcommand{\be}{\begin{equation}}
\newcommand{\ee}{\end{equation}}
\newcommand\half{\textstyle\frac{1}{2}}

\title{Superconducting strings in the two-Higgs doublet model}
\author{Richard A. Battye and Steven J. Cotterill}
\address{Jodrell Bank Centre for Astrophysics, Department of Physics and Astronomy, University of Manchester, Oxford Road, Manchester M13 9PL}

\begin{document}

\begin{abstract}
We present superconducting vortex solutions in the two-Higgs doublet model which has a gauged $U(1)$ Higgs-family symmetry. We write down an ansatz for the solution and study its basic properties, for the case of both global and gauged symmetries. We demonstrate its prima-facie stability using 3D numerical simulations of a global version of the theory, observing the flow of current along the string. We discuss how generic such a phenomena might be and the possible consequences if such a model is found in nature. 
\end{abstract}

\maketitle

\section{Introduction}
    \label{sec:intro}
Topological defects can form during the evolution of the Universe due to the spontaneous breakdown of a symmetry when the vacuum manifold, ${\cal M}$, contains non-contractible loops (see recent reviews~\cite{V&Sbook,Hindmarsh:1994re,2010kdw..book.....V,2015SchpJ..1031682V}). This requires the homotopy group $\pi_1({\cal M})\ne I$ and simplest case is due to the breaking of a $U(1)$ symmetry, for example, the Nielsen-Olesen vortex in the Abelian Higgs-model~\cite{Nielsen:1973cs}.

Witten pointed out that if there is an additional unbroken $U(1)$ symmetry - which would have a conserved Noether  currents and charge - the resulting vortices could be superconducting~\cite{Witten1985,Davis1988a,Davis1988b}. The corresponding currents could lead to many interesting phenomena such as stable loops, known as Vortons~\cite{Davis1989,Brandenberger1996,Lemperiere2003b,Battye2009a,Garaud2013,PhysRevLett.127.241601} and have also been linked with a number of astrophysical phenomena. Some have suggested that these currents are a generic feature~\cite{Davis:1995kk}, but at least within $U(1)\times U(1)$ model usually studied even the phenomena of vortex superconductivity is far from given; it is only possible for specific parameters. In this {\it letter} we will address the question whether it is possible to find superconducting vortex solutions in an extension of the Standard Model (SM) of particle physics, known as the two-Higgs doublet model (2HDM).

As its name suggests this model has two-Higgs doublets, $\Phi_1$ and $\Phi_2$, leading to 5 Higgs particles: the CP-even, $h$ and $H$, CP-odd, $A$ and charged $H^{\pm}$ with the extra particles typically being required to have masses above that of the Higgs boson that has already been detected. Most importantly for the discussion here is that there can be Higgs family symmetries and, in particular, a $U(1)$ symmetry, often denoted $U(1)_{\rm PQ}$ since it can be used to connect Peccei-Quinn (PQ) symmetry to the SM. If this symmetry is a global symmetry then the CP-odd scalar particle is a massless Goldstone boson, $M_{\rm A}=0$, which could be problematic since such a particle would be produced copiously in particle interactions at accelerators. However, it is possible to gauge the symmetry in which case the new gauge field ``eats" the Goldstone boson and acquires a mass via the Higgs-mechanism, allowing the model to become phenomenological viable.

This new $U(1)_{\rm PQ}$ symmetry along with the hypercharge symmetry of the SM, $U(1)_{\rm Y}$, implies that the overall symmetry breaking is $SU(2)_{\rm L}\times U(1)_{\rm Y}\times U(1)_{\rm PQ}\rightarrow U(1)_{\rm EM}$ - where Y stands for hypercharge and EM for ElectroMagnetism -  with vacuum manifold ${\cal M}=S^3\times S^1$~\cite{Brawn2011} such that $\pi_1({\cal M})=\mathbb{Z}$. The associated topological index is the winding number of the vortices and when embedded in 3D these are cosmic strings. Since there is an unbroken $U(1)$ symmetry, it is natural to ask whether these can be superconducting. Moreover, the fact that the unbroken symmetry is EM, the associated current will be the Bosonic electromagnetic current.

In principle the 2HDM can have other discrete or continuous symmetries and a systematic study found the defect solutions associated with global symmetries~\cite{Brawn2011} assuming a neutral vacuum throughout space - corresponding to a massless photon. A number of other authors have considered defect solutions in the 2HDM~\cite{Eto:2018hhg,Eto:2018tnk,Eto:2019hhf,Eto:2020hjb,Eto:2021dca}.

Interestingly, numerical simulations of a global version of the 2HDM (i.e one with no SM gauge fields) starting with random field configurations have been performed~\cite{Viatic2020,BATTYE2023138091,SHF} and they suggest that the neutral vacuum condition is violated in the core of the defects (this has been seen for domain walls, vortices and monopoles) which can have consequences~\cite{Battye:2021dyq}. In addition we have seen that in the context of monopoles and vortices there can be a ``Spontaneous Hopf Fibration"~\cite{SHF} of the $S^3$ to be locally $S^3\cong S^2\times S^1$ with the $S^2$ coupling directly to the $S^2$ associated with the Higgs family symmetry in the case of monopoles and an equivalent phenomena coupling together the two $S^{1}$ parts of the vacuum manifold in the context of vortices.

In this paper we will show that one can naturally also accommodate a superconducting vortex solution in this model extending the vortex solution presented in \cite{SHF} and that it seems to be possible for a relatively large range of parameters. In section~\ref{sec:2HDM} we will introduce the model and in section~\ref{sec:solution} we will present cylindrically symmetric superconducting string solution which can carry both charge and current. In section~\ref{section: sims} we will demonstrate a prima-facie stability of these string solutions in the global version of the 2HDM in which there are no gauge fields. 

\section{Gauged Two-Higgs doublet model with $\text{U}(1)$ Higgs family symmetry}
\label{sec:2HDM}

If we write $\Phi^T=(\Phi_1^T \Phi_2^T)$ then the Lagrangian of the gauged 2HDM with Higgs Family symmetries can be written  as

\be
{\cal L}=(D^{\mu}\Phi)^{\dag}D_{\mu}\Phi-V(\Phi)-\frac{1}{4}W_{\mu\nu}^aW^{a\mu\nu}-\frac{1}{4}Y_{\mu\nu}Y^{\mu\nu}-\frac{1}{4}V_{\mu\nu} V^{\mu\nu}\,,
\ee
where the covariant derivative can be written as 
\be
D_\mu\Phi=\left[(\sigma^0\otimes\sigma^0)\partial_\mu+\half ig(\sigma^0\otimes\sigma^a)W_\mu^a+\half ig^{\prime}(\sigma^0\otimes\sigma^0)Y_\mu+\half i g^{\prime\prime} Q V_{\mu}\right]\Phi\,,
\ee
where $g^{\prime\prime}$ is the new coupling constant, $V_{\mu}$ is the new gauge field associated with $U(1)_{\rm PQ}$ symmetry, $\sigma^0$ is the $2\times 2$ identity matrix and the matrix $Q$ defines the charges of the Higgs fields under the $U(1)_{PQ}$ symmetry. If the charges of $\Phi_i$ are $q_i$ then $Q=\half(q_1+q_2)\sigma^0\otimes\sigma^0+\half(q_1-q_2)\sigma^3\otimes\sigma^0$. The field strength tensors for the gauge fields are $W^{a}_{\mu\nu}=\partial_\mu W^a_{\nu}-\partial_\nu W^a_{\mu}-g\varepsilon^{abc}W^b_{\mu}W^c_{\nu}$, $Y_{\mu\nu}=\partial_\mu Y_\nu-\partial_\nu Y_\mu$ and $V_{\mu\nu}=\partial_\mu V_{\nu}-\partial_\nu V_{\mu}$. 

In this paper we will write the potential in terms of the masses of the Higgs particles, $(M_h,M_H,M_A,M_{H^\pm})$ and two angles: $\beta$ which parameterizes the ratio of the vacuum expectation values (VEVs) of the two Higgs fields, $v_i$ respectively, such that $\tan\beta=v_2/v_1$, and the CP even mixing angle $\alpha$. The two angles being equal,  $\alpha=\beta$, is known as the alignment limit where $h$ has the properties of the SM Higgs; this limit is preferred by a wide range of experimental data. If we write $s_x=\sin x$ and $c_x=\cos x$ then the potential,
\begin{align}
    V = -\mu_1^2\Phi_1^\dagger\Phi_1 - \mu_2^2\Phi_2^\dagger\Phi_2 + \lambda_1(\Phi_1^\dagger\Phi_1)^2 + \lambda_2(\Phi_2^\dagger\Phi_2)^2 + \lambda_3(\Phi_1^\dagger\Phi_1)(\Phi_2^\dagger\Phi_2) + \lambda_4(\Phi_1^\dagger\Phi_2)(\Phi_2^\dagger\Phi_2)\,,
\end{align}
can be expressed as~\cite{Viatic2020}
\begin{align}
V&=-{1\over 2}\left[M_h^2c_\alpha^2+M_H^2s_\alpha^2+(M_h^2-M_H^2)\tan\beta c_\alpha s_\alpha\right]\Phi_1^{\dag}\Phi_1 \nonumber \\
&-{1\over 2}\left[M_h^2s_\alpha^2+M_H^2c_\alpha^2+(M_h^2-M_H^2)\cot\beta c_\alpha s_\alpha\right]\Phi_2^{\dag}\Phi_2 \nonumber \\
&+\left({M_h^2c_\alpha^2+M_H^2s_\alpha^2\over 2c_\beta^2v_{\rm SM}^2}\right)(\Phi_1^{\dag}\Phi_1)^2+\left({M_h^2s_\alpha^2+M_H^2c_\alpha^2\over 2s_\beta^2v_{\rm SM}^2}\right)(\Phi_2^{\dag}\Phi_2)^2 \nonumber \\
&+\left({(M_h^2-M_H^2)c_\alpha s_\alpha+2M_{H^\pm}^2s_\beta c_\beta\over c_\beta s_\beta v_{\rm SM}^2}\right)(\Phi_1^{\dag}\Phi_1)(\Phi_2^{\dag}\Phi_2)-{2M_{H^{\pm}}^2\over v_{\rm SM}^2}(\Phi_1^\dagger\Phi_2)(\Phi_2^\dagger\Phi_1)\,,
\label{potential}
\end{align}
where $v_{\rm SM}=\sqrt{v_1^2+v_2^2}=246\,{\rm GeV}$ is the standard model VEV. We note that we have assumed $M_A=0$ throughout and if in addition $M_H=0$ and $\alpha=\beta=\pi/4$, then the symmetry of the potential is enhanced to $SO(3)_{\rm HF}$ - in what follows we exclude this possibility since in this case the natural defect solutions are monopoles \cite{BATTYE2023138091,SHF}.

\subsection{Mass spectrum}

We can redefine the hypercharge gauge field and coupling with $\Tilde{Y}_\mu = (g^\prime Y_\mu + \half(q_1+q_2)V_\mu^3)/\Tilde{g}^\prime$ and $\Tilde{g}^{\prime 2} = g^{\prime 2} + \frac{1}{4}(q_1+q_2)^2g^{\prime\prime 2}$ as well as defining $\Tilde{g}^{\prime\prime} = \half(q_1-q_2)g^{\prime\prime}$ so that we can write the covariant derivative as
\begin{equation}
    D_\mu\Phi = [(\sigma^0\otimes\sigma^0)\partial_\mu + \half ig(\sigma^0\otimes\sigma^a)W_\mu^a + \half i\Tilde{g}^\prime(\sigma^0\otimes\sigma^0)\Tilde{Y}_\mu + \half i\Tilde{g}^{\prime\prime}(\sigma^3\otimes\sigma^0)V_\mu]\Phi \,, \label{eq: Covar deriv}
\end{equation}
which is beneficial because these terms will enter into the masses of the gauge particles in the same way. In a neutral vacuum state we have that $\Phi_1^\dagger = \frac{v_{\text{SM}}}{\sqrt{2}}\begin{pmatrix}0 & \cos\beta \end{pmatrix}$, $\Phi_2^\dagger = \frac{v_{\text{SM}}}{\sqrt{2}}\begin{pmatrix}0 & \sin\beta \end{pmatrix}$ and spatial derivatives are zero so that
\begin{eqnarray}
{|D_i\Phi|^2\over v_{\rm SM}^2}=\textstyle\frac{1}{8}\left(g^2W_i^a W_i^a+\Tilde{g}^{\prime 2}\Tilde{Y}_i \Tilde{Y}_i+\Tilde{g}^{\prime\prime 2}\Tilde{V}_i \Tilde{V}_i-2g\Tilde{g}^{\prime}\Tilde{Y}_i W_i^a{\hat z}^a\right)+\textstyle\frac{1}{4}\cos 2\beta\left[\Tilde{g}^{\prime}\Tilde{g}^{\prime\prime}\Tilde{Y}_i \Tilde{V}_i-g\Tilde{g}^{\prime\prime}\Tilde{V}_iW_i^a{\hat z}^a\right]\,.
\end{eqnarray}
If we define ${\cal P}^{ab}=\delta^{ab}-{\hat z}^a{\hat z}^b$ and $Z_i = (\Tilde{g}'\Tilde{Y}_i-gW_i^a\hat{z}^a)/\Tilde{g}$ (the Z boson in the SM), where $\Tilde{g} = \sqrt{g^2+\Tilde{g}'^2}$, then we can write 
\begin{equation}
|D_i\Phi|^2=\textstyle\frac{1}{2}v_{\rm SM}^2\bigg\{\textstyle\frac{1}{4}\left[g^2{\cal P}^{ab}W_i^a W_i^b+\Tilde{g}^{\prime\prime 2}\Tilde{V}_i \Tilde{V}_i + \Tilde{g}^2Z_i Z_i\right] +\textstyle\frac{1}{2}\Tilde{g}g''Z_i V_i\cos 2\beta\bigg\}\,.
\end{equation}
There is a massless particle, the photon, that corresponds to the degree of freedom perpendicular to $Z_i$, $A_i = (g\Tilde{Y}_i + \Tilde{g}^\prime W_i^a\hat{z}^a)/\Tilde{g}$. There are also the usual, charged weak bosons, $\mathcal{W}_i^\pm = (\mathcal{W}_i^a\hat{x}^a\mp i\mathcal{W}_i^a\hat{y}^a)/\sqrt{2}$ with mass $M_\mathcal{W}^2 = \frac{1}{4}g^2v_{\text{SM}}^2$. The masses of the remaining two (neutral) particles are given by the eigenvalues of the mass matrix
\begin{equation}
    \frac{1}{8}v_{\text{SM}}^2\begin{pmatrix}Z_i & \Tilde{V}_i\end{pmatrix}\begin{pmatrix}
        \Tilde{g}^2 & \Tilde{g}\Tilde{g}''\cos 2\beta \\
        \Tilde{g}\Tilde{g}''\cos 2\beta & \Tilde{g}''^2
    \end{pmatrix}\begin{pmatrix}Z_i \\ \Tilde{V}_i\end{pmatrix}\,,
\end{equation}
which are $\lambda_\pm = \frac{1}{2}(\Tilde{g}^2+\Tilde{g}^{\prime\prime 2}) \pm \frac{1}{2}\sqrt{(\Tilde{g}^2+\Tilde{g}^{\prime\prime 2})^2 - 4\Tilde{g}^2\Tilde{g}^{\prime\prime 2}\sin^2 2\beta}$ and the masses are $M^2_\pm={1\over 4}v_{\rm SM}^2\lambda_\pm$.

\subsection{Bilinear forms}

It has been shown~\cite{Ivanov2007,Brawn2011,BATTYE2023138091} that the eight degrees of freedom of the Higgs field $\Phi$ can be encoded in terms of bilinear forms defined by $n^{a}=-\Phi^{\dag}(\sigma^0\otimes\sigma^a)\Phi$ and $R^A=(R^{\mu},R^4,R^5)$ where $R^{\mu}=\Phi^{\dag}(\sigma^\mu\otimes\sigma^0)\Phi$ and ${\tilde R}=R^4+iR^5=2\Phi_1^{T}i\sigma^2\Phi_2$. The vector $n^{a}$ transforms under the SM model degrees of freedom associated with $SU(2)_{\rm L}$, the complex scalar ${\tilde R}$ encodes the hypercharge degrees of freedom, $U(1)_{\rm Y}$ and, by virtue of the fact that the potential of any 2HDM that respects the SM symmetries can be written completely in terms of $R^\mu$, this represents the new Higgs family components (and the total magnitude of the field). 

The potential that we have chosen in equation (\ref{potential}) is constructed so that the only dependence on $R^1$ and $R^2$ is in the combination $(R^1)^2+(R^2)^2$, such that the potential is symmetric under rotations between $R^1$ and $R^2$, which is the $U(1)_{\text{PQ}}$ symmetry. Any closed path which has a non-zero winding number associated with this rotation will contain a string somewhere inside, which is identified by $(R^1)^2+(R^2)^2=0$. These strings often exhibit additional structure, such as imprints of the non-trivial topology of $R^\mu$ upon $n^a$ and local non-neutrality (and therefore a locally massive photon) if $R_\mu R^\mu \neq 0$ in the core \cite{SHF}.

\section{Global superconducting string}
\label{sec:solution}

The global vortex solution~\cite{SHF} with winding number $n$ can be expressed as 
\begin{equation}
   \Phi(r,\theta)=\frac{v_{\text{SM}}}{\sqrt{2}}\begin{pmatrix}e^{-\half in\theta}& 0\\0 & e^{\half in\theta}\end{pmatrix}\otimes \left[\begin{pmatrix}e^{-\half in\theta}& 0\\0 & e^{\half in\theta}\end{pmatrix}
   \begin{pmatrix}\cos\half\gamma & \sin\half\gamma\\-\sin\half\gamma & \cos\half\gamma\end{pmatrix}\right]\begin{pmatrix}0\\ f_1\\f_+\\f_2\end{pmatrix}\,
   \,, \label{eq: no current string}
\end{equation}
where $f_1=f_1(r)$, $f_2=f_2(r)$, $f_+=f_+(r)$ and $\gamma=\gamma(r)$. The vacuum expectation values of the 2HDM fields are often parameterised as $\langle\Phi^\dagger\rangle = \frac{v_{\rm SM}}{\sqrt{2}}\begin{pmatrix}0 & v_1 & v_+ & v_2e^{i\xi}\end{pmatrix}$ and our ansatz is constructed by acting on this with elements of the symmetry groups. The functions $f_a$ approach the corresponding vacuum values, $v_a$, far away from the string and the final vacuum parameter $\xi$ is absorbed into the $\text{U(1)}_{\rm PQ}$ symmetry. The $\rm SU(2)_{\rm L}$ matrix in the square brackets uses two distinct degrees of freedom, with the left-hand matrix corresponding to the spontaneous Hopf fibration \cite{SHF} and the right-hand matrix being sourced by the current, $f_+\partial_\mu f_2 - f_2\partial_\mu f_+$.

In order to create a current-carrying superconducting string, it is useful to introduce the four-vector $\psi_\mu = \omega\hat{t}_\mu - k\hat{z}_\mu$ which points in the space-time direction of the current. We will sometimes choose to normalise this vector with $\psi_\mu = \psi\hat{\psi}_\mu$, where $\psi$ is a positive real number and $\hat{\psi}_\mu\hat{\psi}^\mu = \pm 1$ depending on whether the current is space-like or time-like. Note that the magnitude is $\psi_\mu\psi^\mu = \omega^2-k^2\equiv \kappa$, which is familiar from the original example of a bosonic superconducting string \cite{Witten1985}. Strings with time-like currents ($\kappa>0$) are known as electric, strings with space-like currents ($\kappa<0$) are known as magnetic and those with null currents ($\kappa=0$) are known as chiral. This nomenclature is in reference to gauged superconducting strings, where the associated electromagnetic field has the property that the Lorentz invariant quantity, $|\mathbf{E}|^2-|\mathbf{B}|^2$, has the same sign as $\kappa$.


In order for the currents to be localised to the string, we need to use the degree of freedom that is unbroken in the vacuum (corresponding to the photon) and act on Eq. (\ref{eq: no current string}) with $e^{\frac{1}{2}i\psi_\mu x^\mu}[\sigma^0 \otimes e^{\frac{1}{2}i\psi_\nu x^\nu\sigma^3}]$ to obtain the ansatz for a superconducting string
\begin{equation}
    \Phi = \frac{v_{\text{SM}}}{\sqrt{2}}\begin{pmatrix}
        g_1(r)e^{-in\theta}e^{i(\omega t + kz)} \\
        g_2(r) \\
        g_3(r)e^{i(\omega t + kz)} \\
        g_4(r)e^{in\theta}
    \end{pmatrix}\,,
\end{equation}
 where we have replaced $g_1 = f_1\sin\frac{1}{2}\gamma$, $g_2 = f_1\cos\frac{1}{2}\gamma$, $g_3 = f_+\cos\frac{1}{2}\gamma + f_2\sin\frac{1}{2}\gamma$ and $g_4 = f_2\cos\frac{1}{2}\gamma - f_+\sin\frac{1}{2}\gamma$, as this makes it easier to calculate the solutions numerically. The current-carrying phases are only attached to the parts of the field that are zero in the vacuum, confirming that the current is confined to the string.

Under this ansatz the equations of motion are
\begin{align}
    \frac{d^2g_1}{dr^2} + \frac{1}{r}\frac{dg_1}{dr} - \bigg[ \bigg(\frac{n}{r}\bigg)^2 + \lambda_1(g_1^2+g_2^2-\eta_1^2) + \frac{1}{2}\lambda_3(g_3^2+g_4^2) - \kappa \bigg]g_1 \nonumber \\
    - \frac{1}{2}\lambda_4(g_1g_3+g_2g_4)g_3 &= 0 \,, \label{eq: g1 eom}\\
    \frac{d^2g_2}{dr^2} + \frac{1}{r}\frac{dg_2}{dr} - \bigg[\lambda_1(g_1^2+g_2^2-\eta_1^2) + \frac{1}{2}\lambda_3(g_3^2+g_4^2) \bigg]g_2 - \frac{1}{2}\lambda_4(g_1g_3+g_2g_4)g_4 &= 0 \,, \label{eq: g2 eom}\\
    \frac{d^2g_3}{dr^2} + \frac{1}{r}\frac{dg_3}{dr} - \bigg[\lambda_2(g_3^2+g_4^2-\eta_2^2) + \frac{1}{2}\lambda_3(g_1^2+g_2^2) - \kappa \bigg]g_3 - \frac{1}{2}\lambda_4(g_1g_3+g_2g_4)g_1 &= 0 \,, \label{eq: g3 eom}\\
    \frac{d^2g_4}{dr^2} + \frac{1}{r}\frac{dg_4}{dr} - \bigg[ \bigg(\frac{n}{r}\bigg)^2 + \lambda_2(g_3^2+g_4^2-\eta_2^2) + \frac{1}{2}\lambda_3(g_1^2+g_2^2) \bigg]g_4 - \frac{1}{2}\lambda_4(g_1g_3+g_2g_4)g_2 &= 0 \label{eq: g4 eom}\,,
\end{align}
where we have defined $\eta^2_a = \mu_a^2/\lambda_av_{\rm SM}^2$ and omitted factors of $v_{\rm SM}$, as we will shortly argue that it can be set to $1$ without loss of generality.

\subsection{Numerical solutions and properties}

Using a grid spacing of $\Delta r = 0.01$ and $r_{\text{max}}=100$, we solve equations (\ref{eq: g1 eom}) - (\ref{eq: g4 eom}) numerically, with boundary conditions $g_1(0)=g_2^\prime(0)=g_3^\prime(0)=g_4(0)=0$, $g_1(r_{\text{max}})=g_3(r_{\text{max}})=0$, $g_2(r_{\text{max}})=\cos\beta$ and $g_4(r_{\text{max}})=\sin\beta$. It is also necessary to set the value of $\kappa$ in the equations of motion, for which there are some technical details to discuss. In the magnetic regime or chiral regimes ($\kappa\leq 0)$, it is sufficient to fix the value of $\kappa$ directly, but in the electric regime it is better to fix the value of the charge per unit length instead, $q = \pi\omega v_{\text{SM}}^2\int rdr(g_1^2 + g_3^2)$. The reason for the two different approaches is that $\kappa$ and $q$ are the physically conserved quantities in the two regimes. The quantity $\kappa$ directly corresponds to a conserved winding number in the magnetic regime, while in the electric regime it is the Noether charge that is the important conserved quantity \cite{2022JHEP...04..005B}. In order to keep $q$ fixed, we arbitrarily choose to work in the frame where $k=0$, which is always achievable in the electric regime by performing a Lorentz boost directed along the string. In this frame, $\kappa = \omega^2$, therefore we can numerically evaluate the integral and update the value of $\kappa$ at every iteration so that $q$ remains constant.

We choose to set the model parameters so that we satisfy the alignment limit, $\alpha=\beta$, as well as $\tan\beta=1$, $\delta\equiv M_H/M_h = 2$ and $\epsilon\equiv M_{H\pm}/M_h=1$. Given that these values are fixed, we can set the mass of the Higgs boson, $M_h$, and the SM VEV, $v_{\text{SM}}$, to any value without changing the physics, it simply corresponds to a rescaling of lengths and field magnitudes. We will, therefore, assume that $M_h=v_\text{SM}=1$ from here on.

\begin{figure}[!t]
    \centering
    \subfloat[$\kappa=-0.015$]{
        \centering
        \includegraphics[trim={0.1cm 0cm 1.3cm 1.3cm},clip,width=0.48\linewidth]{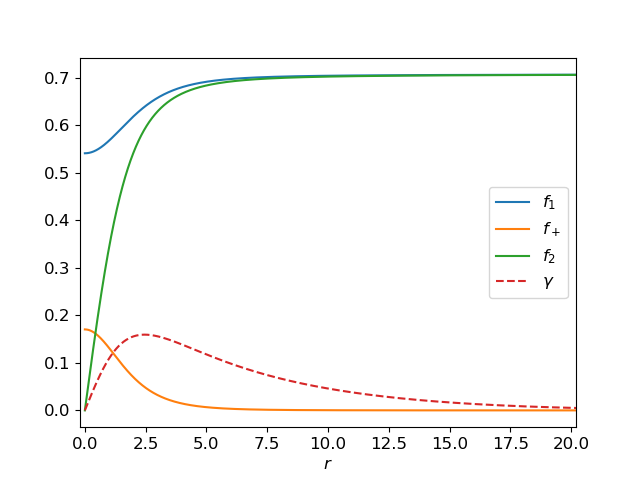}
    }\hfill
    \subfloat[$q=1$ ($\kappa\approx 2\times10^{-3}$)]{
        \centering
        \includegraphics[trim={0.1cm 0cm 1.3cm 1.3cm},clip,width=0.48\linewidth]{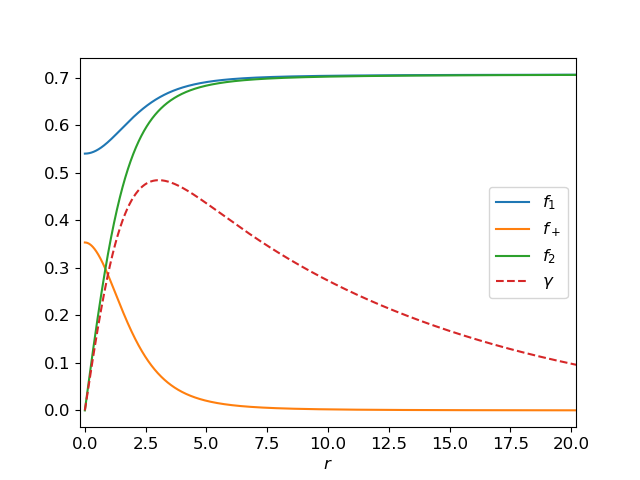}
    }\hfill
    \caption{Global superconducting string solutions in the 2HDM in the alignment limit with $\tan\beta=1$, $\delta=2$ and $\epsilon=1$, clearly showing condensation at the core of the string which is enhanced for larger values of $\kappa$.}
    \label{fig: superconducting string solutions}
\end{figure}

In Figure \ref{fig: superconducting string solutions} we show two examples of current-carrying string solutions with the same set of model parameters, but one is magnetic and the other is electric. Here, we see the condensation of $g_1$ and $g_3$ onto the string, as well as reduced or enhanced condensation in the magnetic or electric regimes, respectively. This is a typical feature of current-carrying strings that is caused by the contribution of $\kappa$ to the effective mass of the condensates \cite{V&Sbook}. We also note that, since $R_\mu R^\mu = f_1^2f_+^2 = (g_2g_3-g_1g_4)^2$ is clearly non-zero at the centre of the string, the photon gains a mass in the core of the string and it will gain a larger mass in the electric regime than in the magnetic.

\subsection{Simulations} \label{section: sims}

As a confirmation that these solutions are correct, and that they are stable, we have performed 3D simulations of a few of the solutions. We can construct a $z-$directed string by using the 1D profile functions, that we have numerically calculated, to set the magnitudes of the field values, while the phase is assigned to the relevant components by both the angle in the $x-y$ plane (for the vortex winding) and using $z$ and $t$ (for the current). The simulations that we performed are run on a 3D grid with $256$ points in each dimension, a lattice spacing of $\Delta x = 0.25$ (so that the physical side length of the box and length of the string is $L=64$) and timesteps of $\Delta t = 0.1$. We used periodic boundary conditions in the $z$-direction and fixed boundary conditions in the $x$ and $y$ directions. This is a small simulation compared to modern network simulations of strings, but this was done in order to allow us to run them for a longer period of time, which is naturally a more demanding test of stability. 

\begin{figure}[!t]
    \centering
    \subfloat[$t=0$.]{
        \centering
        \includegraphics[trim={10cm 8cm 10cm 8cm},clip,width=0.3\linewidth]{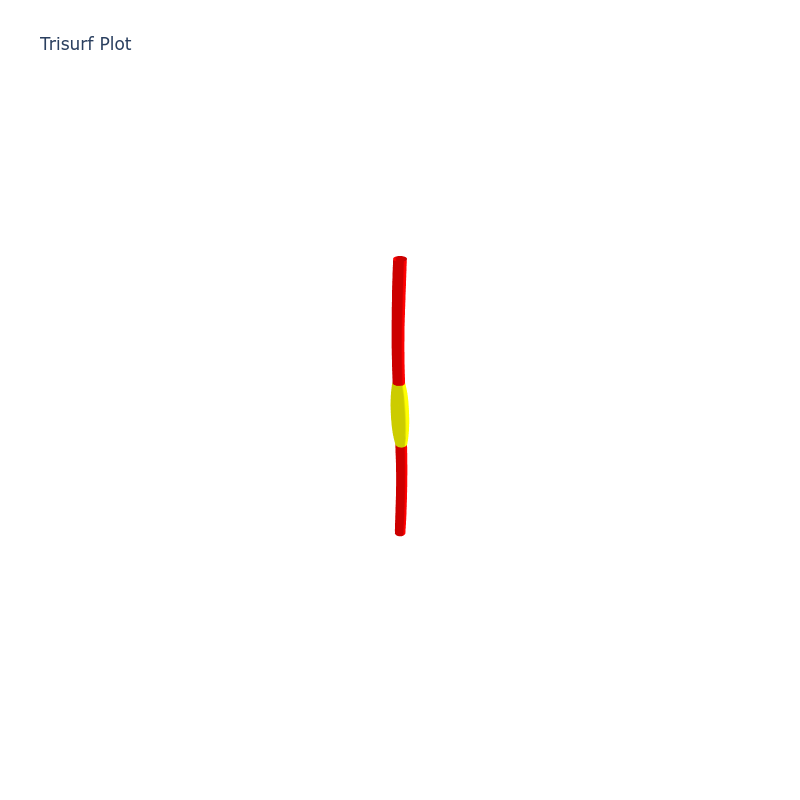}
    }\hfill
    \subfloat[$t=14$.]{
        \centering
        \includegraphics[trim={10cm 8cm 10cm 8cm},clip,width=0.3\linewidth]{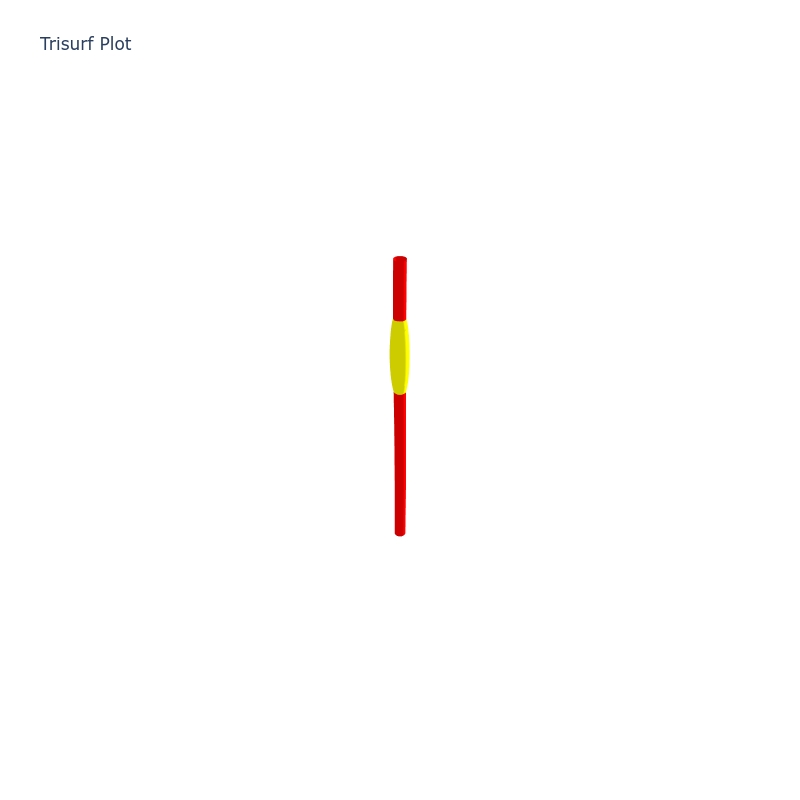}
    }\hfill
    \subfloat[$t=28$.]{\label{fig: variable xi}
        \centering
        \includegraphics[trim={10cm 8cm 10cm 8cm},clip,width=0.3\linewidth]{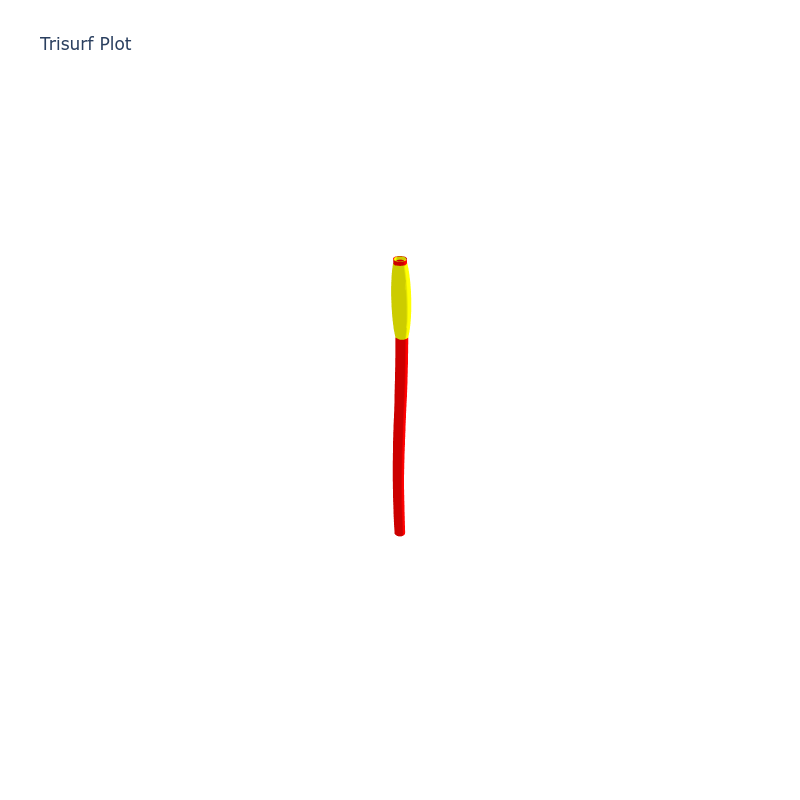}
    }\hfill
    \caption{Snapshots from the evolution of an approximately chiral string, with $N=1$ and $\omega=9.81\times 10^{-2}$, that has been sinusoidally perturbed. The red isosurfaces represent locations where $R_1^2+R_2^2=0.225$, which is 90\% of the vacuum value for this quantity ($0.25$). The yellow isosurfaces are where $R_4 = 0.1$, which has a vacuum value of zero. The full simulation runs for long enough that the current completes around 60 full loops without any signs of instability, lasting for $140$ light crossing times by the end of the simulation.}
    \label{fig: chiral perturb sim}
\end{figure}

In Figure \ref{fig: chiral perturb sim} we show isosurfaces of $R_1^2+R_2^2=0.225$ in red and $R_4=0.1$ (which is equal to $-v_{\text{SM}}^2f_1f_+\cos\psi_\mu x^\mu$ for our ansatz) in yellow at three different times during the evolution of an approximately chiral current-carrying string. The string is initialised with a phase frequency of $\omega = 9.81\times 10^{-2}$ and a winding number of $N=1$ (note that $k=2\pi N/L$). We have also added a slight sinusoidal shift in the position of the string along the $z$ direction, with an amplitude of $0.5$, for a more interesting test of the stability of the solution.

The simulation clearly shows the expected behaviour of the current travelling along the string and, due to the applied perturbation, an additional oscillation of the entire object. Numerical simulations can never prove that a field configuration is absolutely stable, they can only place a lower limit on its lifetime, but we have shown that this object lives for a long period of time, with the current completing more than $60$ full loops and no signs of instability by the end of the simulation, which is over $140$ light-crossing times. 

\section{Gauged superconducting string}

In the gauged case, we take the approach of working in the unitary gauge where we set $\Phi \to \Bar{\Phi} = \frac{v_{\text{SM}}}{\sqrt{2}}\begin{pmatrix} 0 & f_1 & f_+ & f_2 \end{pmatrix}$ and $(W^a_\mu,\,Y_\mu,\,V_\mu) \to (\mathcal{W}^a_\mu,\,\mathcal{Y}_\mu,\,\mathcal{V}_\mu)$ via the local symmetry transformation $U_Y(U_H\otimes U_L)$. Under this transformation, the respective gauge fields transform like $\mathcal{W}_\mu^a = \frac{1}{g}w_\mu^a + W_\mu^b\mathcal{R}^{ba}_L$, $\mathcal{Y}_\mu = \frac{1}{g^\prime}y_\mu + Y_\mu$ and $\mathcal{V}_\mu = \frac{1}{g^{\prime\prime}}v_\mu + V_\mu$ where $U_L^\dagger\sigma^a U_L = \mathcal{R}^{ab}_L\sigma^b$, $U_L^\dagger\partial_\mu U_L = \frac{1}{2}iw_\mu^a\sigma^a$, $U_Y^\dagger\partial_\mu U_Y = \frac{1}{2}iy_\mu\sigma^0$ and $U_H^\dagger\partial_\mu U_H = \frac{1}{2}iv_\mu\sigma^3$. In this gauge, superconductivity will be represented by non-zero terms in the $\mathcal{Y}_\mu$ and $\mathcal{W}_\mu^3$ fields and the equations of motion are
\begin{align}
    \Bar{D}_\mu\Bar{D}^\mu\Bar{\Phi} + \frac{\partial V}{\partial\Bar{\Phi}^\dagger} &= 0\,, \quad\quad \partial_\mu\mathcal{W}^{a\mu\nu} - g\epsilon^{abc}\mathcal{W}_\mu^b\mathcal{W}^{c\mu\nu} = J^{a\nu}_W\,,
\end{align}
$\partial_\mu\mathcal{Y}^{\mu\nu}=J^\nu_Y$ and $\partial_\mu\mathcal{V}^{\mu\nu}=J^\nu_V$ where $\Bar{D}_\mu$ is the covariant derivative defined in \eqref{eq: Covar deriv}, but with the tildes neglected and using the unitary gauge versions of the gauge fields. The currents associated with each symmetry group are
\begin{align}
    J^{a\nu}_W &= \frac{1}{2}ig\bigg[ \Bar{\Phi}^\dagger(\sigma^0\otimes\sigma^a)(\Bar{D}^\nu\Phi) - (\Bar{D}^\nu\Phi)^\dagger(\sigma^0\otimes\sigma^a)\Phi \bigg] \,, \\
    J^\nu_Y &= \frac{1}{2}ig^\prime\bigg[ \Bar{\Phi}^\dagger(\Bar{D}^\nu\Phi) - (\Bar{D}^\nu\Phi)^\dagger\Phi \bigg] \,, \\
    J^\nu_V &= \frac{1}{2}ig^{\prime\prime}\bigg[ \Bar{\Phi}^\dagger(\sigma^3\otimes\sigma^0)(\Bar{D}^\nu\Phi) - (\Bar{D}^\nu\Phi)^\dagger(\sigma^3\otimes\sigma^0)\Phi \bigg] \,.
\end{align}
The electromagnetic current is the combination $\frac{1}{2}(g^\prime J^{3\nu}_W+ gJ^\nu_Y)$, which will become important for finding electric string solutions where it is better to fix the associated Noether charge, instead of $\kappa$.

The non-zero gauge field coefficients can be extracted directly from the transformations required in order to fix this gauge when starting from the global ansatz, plus a few additional terms that become non-zero due to couplings between the gauge fields themselves. As we are starting from the global ansatz, the gauge fields are all zero before we make the transformation so we just need to determine which components of $w_\mu^a$, $y_\mu$ and $v_\mu$ are non-zero.

The isospin transformation that was applied to generate the global field configuration was
\begin{equation}
    U_L = \begin{pmatrix}
        \cos\frac{1}{2}\gamma_1e^{\frac{1}{2}i(\psi_\mu x^\mu - n\theta)} & \sin\frac{1}{2}\gamma_1e^{\frac{1}{2}i(\psi_\mu x^\mu - n\theta)} \\
        -\sin\frac{1}{2}\gamma_1e^{-\frac{1}{2}i(\psi_\mu x^\mu - n\theta)} & \cos\frac{1}{2}\gamma_1e^{-\frac{1}{2}i(\psi_\mu x^\mu - n\theta)}
    \end{pmatrix} \,,
\end{equation}
which has
 \begin{equation}
     U_L^\dagger\partial_\mu U_L = \frac{1}{2}i\bigg[ \sin\gamma_1\Big(\psi_\mu - \frac{n}{r}\hat{\theta}_\mu\Big)\hat{x}^a + \frac{d\gamma_1}{dr}\hat{r}_\mu\hat{y}^a + \cos\gamma_1\Big(\psi_\mu - \frac{n}{r}\hat{\theta}_\mu\Big)\hat{z}^a \bigg]\sigma^a \,.
 \end{equation}
There is also a hypercharge transformation, $U_Y = e^{\frac{1}{2}i\psi_\mu x^\mu}$, for which $U_Y^\dagger\partial_\mu U_Y = \frac{1}{2}i\psi_\mu$, and the accidental symmetry transformation, $U_H = \text{diag}(e^{-\frac{1}{2}in\theta}, e^{\frac{1}{2}in\theta})$, for which $U_H^\dagger\partial_\mu U_H = -\frac{1}{2}i\frac{n}{r}\hat{\theta}_\mu\sigma^3$. Therefore, we can fix the gauge at the expense of gaining seven non-zero components of the gauge field, $W_\mu^a = (W_\theta^1\hat{x}^a + W_\theta^3\hat{z}^a)\hat{\theta}_\mu + W_r^2\hat{y}^a\hat{r}_\mu + (W_\psi^1\hat{x}^a + W_\psi^3\hat{z}^a)\hat{\psi}_\mu$, $Y_\mu = Y_\psi\hat{\psi}_\mu$ and $V_\mu = V_\theta\hat{\theta}_\mu$. It should be noted that some of these transformations are singular at the origin which results in seemingly unphysical gauge fields, for example, they can have non-zero components pointing in the $\theta$ direction at the origin. This is not a problem from the point of view of finding the 1D profile functions of the fields and the physical meaning can be restored by reversing the gauge transformations once the solutions are found. We use this gauge primarily as a means to distinguish components which can consistently remain zero. On that note, we will also want to add two additional components, $Y_\theta$ and $V_\psi$, as it is easy to see from the equations of motion that they will become non-zero, in general, due to interactions with other non-zero gauge fields through the currents.

Therefore the complete ansatz for the gauged, current-carrying string can be written as $f_i = f_i(r)$, $gW_\mu^a = \frac{n}{r}[h_1(r)\hat{x}^a + (1-h_3(r))\hat{z}^a]\hat{\theta}_\mu + h_2(r)\hat{y}^a\hat{r}_\mu + \psi[h_{1c}(r)\hat{x}^a + (1-h_{3c}(r))\hat{z}^a]\hat{\psi}_\mu$, $g^\prime Y_\mu = \frac{n}{r}b(r)\hat{\theta}_\mu + \psi(1-b_c(r))\hat{\psi}_\mu$ and $g^{\prime\prime}V_\mu = \frac{n}{r}(1-H(r))\hat{\theta}_\mu + \psi H_c(r)\hat{\psi}_\mu$. These functions must satisfy the boundary conditions that all of the fields are fixed to zero at the centre except for $f_1$, $f_+$ and $h_2$, which have $f_1^\prime(0)=f_+^\prime(0)=h_2^\prime(0)=0$, whereas far from the string they satisfy $f_1(\infty)=\cos\beta$, $f_2(\infty)=\sin\beta$, $h_3(\infty)=H(\infty)=h_{3c}(\infty)=b_c(\infty)=1$ and the rest go to zero.

\begin{figure}[!t]
    \centering
    \subfloat{
        \centering
        \includegraphics[trim={0.1cm 0cm 1.3cm 1.3cm},clip,width=0.48\linewidth]{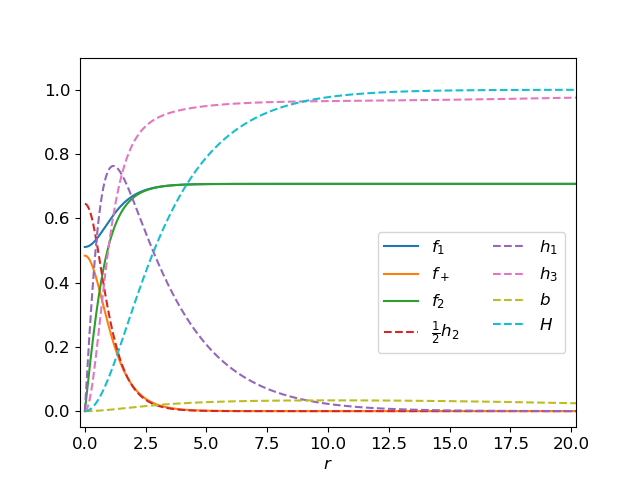}
    }
    \subfloat{
        \centering
        \includegraphics[trim={0.1cm 0cm 1.3cm 1.3cm},clip,width=0.48\linewidth]{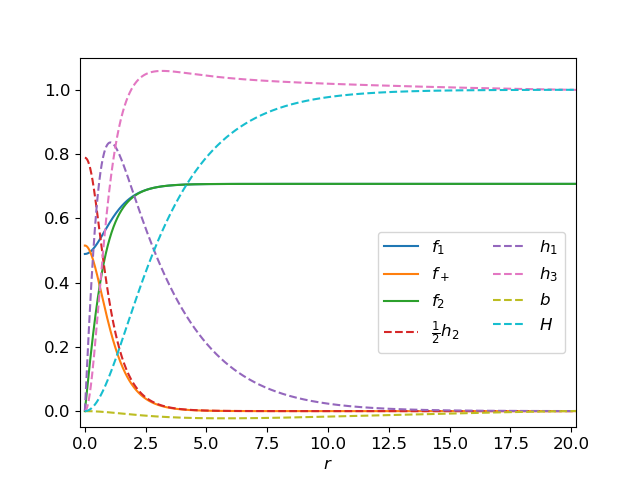}
    }\vspace{-0.7cm}
    \setcounter{subfigure}{0}
    \subfloat[$\kappa=-0.3$]{
        \centering
        \includegraphics[trim={0.1cm 0cm 1.3cm 1.3cm},clip,width=0.48\linewidth]{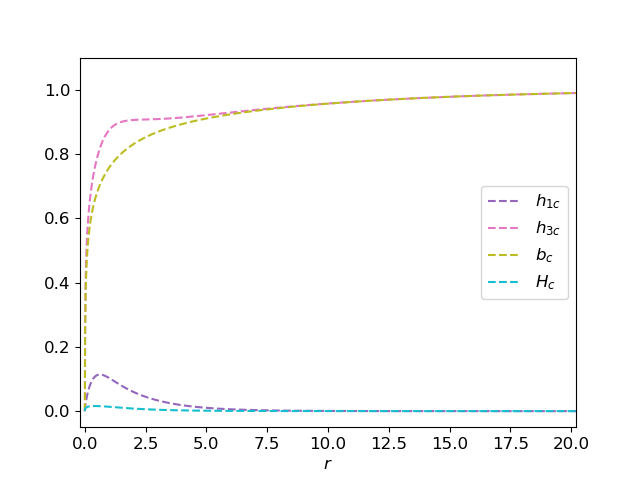}
    }
    \subfloat[$q=0.1$ ($\kappa\approx 0.45$)]{
        \centering
        \includegraphics[trim={0.1cm 0cm 1.3cm 1.3cm},clip,width=0.48\linewidth]{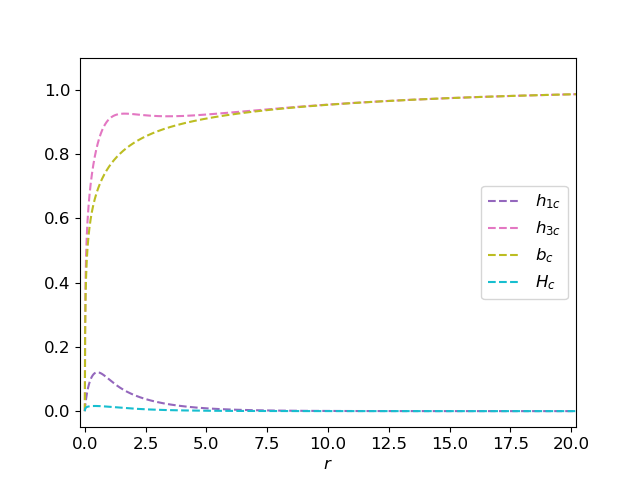}
    }
    \caption{Gauged superconducting string solutions in the 2HDM in the alignment limit with $\tan\beta=1$, $\delta=2$, $\epsilon=1$, $g=g^{\prime\prime}=0.652$ and $g^\prime=0.357$. The left (right) panels show a magnetic (electric) string, with the bottom panels showing the field components that only become non-zero when there is a current on the string and the top panels showing the rest of the non-zero field components. Note that we have halved $h_2$ so that it fits on this scale.}
    \label{fig: gauged superconducting string solutions}
\end{figure}

We solve the static equations of motion numerically, with the same numerical procedure that we used for the global case. In figure \ref{fig: gauged superconducting string solutions} we show a couple of these solutions, one magnetic and one electric, for the same parameters as the solutions shown in figure \ref{fig: superconducting string solutions}, plus the additional parameters $g = 0.652$ and $g^\prime = 0.357$, which are the SM values, as well as choosing $g^{\prime\prime} = g$. For the magnetic solution we have directly fixed $\kappa$, whereas for the electric solution it is the charge per unit length that is kept fixed where, under our ansatz, this can be written as $q = \frac{1}{2}\pi\psi\int rdr f_+^2(2-h_{3c}-b_{c}+H_c)$. Since $\kappa = \psi^2$ in the electric regime, it can be determined numerically, for a given value of $q$, by evaluating this integral. Again, these plots clearly display that the condensation of $f_+$ is enhanced in the electric regime and decreased in the magnetic regime.

\begin{figure}[!t]
    \centering
    \subfloat[$\epsilon=2$]{
        \centering
        \includegraphics[trim={0.1cm 0cm 1.3cm 1.3cm},clip,width=0.48\linewidth]{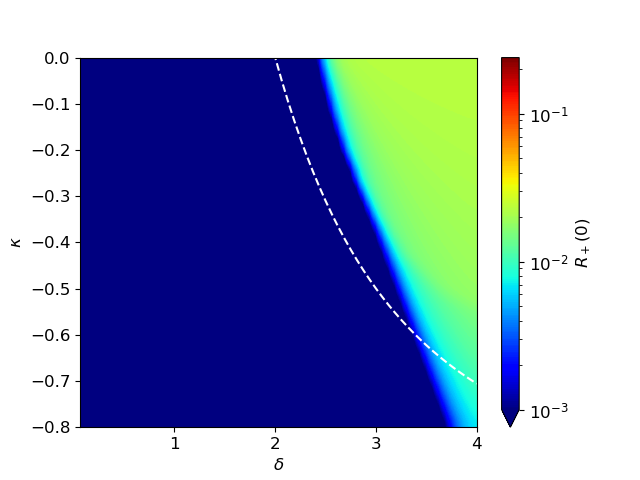}
    }\hfill
    \subfloat[$\delta=2$]{
        \centering
        \includegraphics[trim={0.1cm 0cm 1.3cm 1.3cm},clip,width=0.48\linewidth]{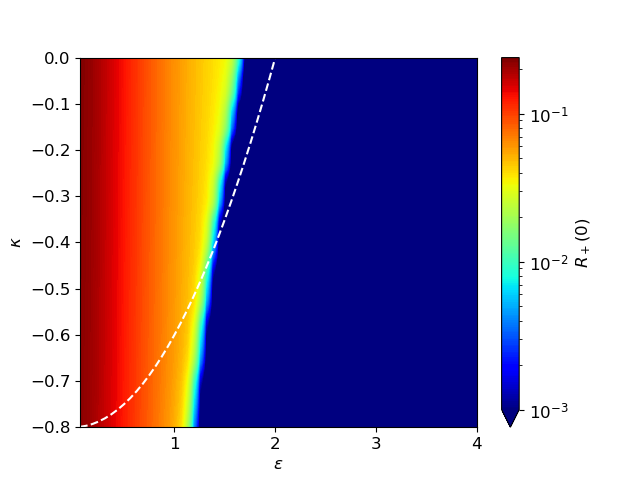}
    }\hfill
    \caption{A contour plot showing the value of $R_+\equiv R_\mu R^\mu$ at the centre of the string, for various values of $\kappa$, $\epsilon$ and $\delta$, with the remaining parameters set to $\alpha=\beta=\pi/4$, $g=g^{\prime\prime} = 0.652$ and $g^\prime = 0.357$. Both plots are generated from a grid of $64^2$ solutions, with parameter values varying linearly within the limits shown. There is a clear, sharp transition line (after which $R_+(0)$ falls quickly to zero) which creates a slightly jagged appearance but this is just an artefact of the numerical resolution.}
    \label{fig: param space}
\end{figure}

In a previous paper \cite{SHF} we used an analysis of the effective mass of $f_+$ about a solution with $f_+=0$ to predict when it would condense onto the cores of the string solutions. If $\alpha=\beta=\pi/4$, as we have set here, then the critical region of the parameter space simplifies to the line $\epsilon = \delta$. However, for a current carrying string, $\kappa$ contributes to the effective mass and this condition generalises to $(1+\delta^2)\kappa_{\text{crit}} = \epsilon^2-\delta^2$. In Figure \ref{fig: param space} we compare this generalised prediction to the results from our numerical solutions by displaying the value of $R_+ \equiv R_\mu R^\mu$ at the centre of the string for a range of different parameter sets and choices of $\kappa$, in the magnetic regime. Despite the fact that this prediction neglects the gradient energy, it continues to provide a good estimate of the critical value.

\begin{figure}[!t]
    \centering
    \includegraphics[trim={0.1cm 0cm 1.3cm 1.3cm},clip,width=0.75\linewidth]{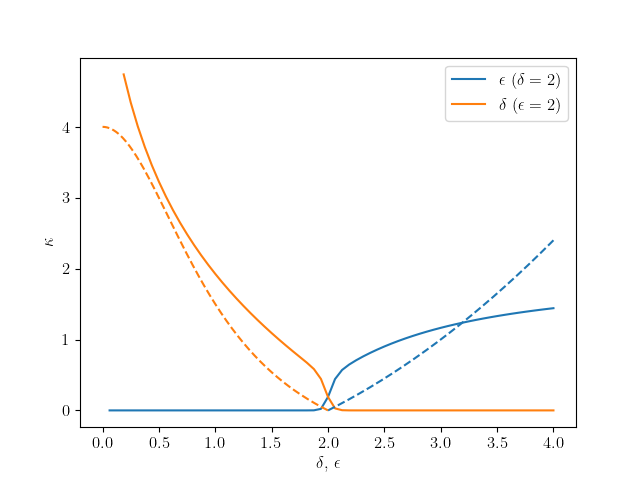}
    \caption{The calculated value for $\kappa$ for a range of electric string solutions with $q=10^{-6}$ (solid lines) compared to our prediction for $\kappa_{\rm crit}$ (dashed lines). The blue lines have variable $\epsilon$ and $\delta=2$ fixed, while the orange lines have variable $\delta$ and $\epsilon=2$ fixed.}
    \label{fig: elec kappa pred}
\end{figure}

In the electric regime, a similar analysis is not as straightforward to perform. This is because, if one keeps $\kappa$ fixed, it has a maximum value (which changes with the parameters) above which $f_+$ will be non-zero in the vacuum. This is an unphysical solution that would have infinite energy. These problems are not encountered if one fixes $q$ instead, as we have advocated throughout this paper. However, this also means that $f_+$ cannot be zero everywhere, unless $q=0$. Despite this, the argument for a critical value of $\kappa$ is still valid, and hence for small $q$ we would expect to see $\kappa\approx\kappa_{\rm crit}$ when $\kappa_{\rm crit}>0$ and very small $\kappa$ otherwise. We have found string solutions with $q=10^{-6}$, over a range of $\epsilon$ and $\delta$, and calculated the resulting value for $\kappa$, which is presented in Fig.~\ref{fig: elec kappa pred}. This clearly shows the behaviour that we expected and, similarly to the magnetic regime, we consider our expression for $\kappa_{\rm crit}$ to provide a good estimate for the true value. For a more accurate prediction in both the electric and magnetic regimes, the numerical method used in \cite{2022JHEP...04..005B} can be adapted for use in the 2HDM.

\section{Conclusion}

In this work, we have shown that current-carrying, superconducting string solutions exist in 2HDMs with a $U(1)_{\text{PQ}}$ symmetry, by calculating the solutions numerically in both the global and gauged cases. More importantly, there is some evidence to suggest that these strings should be expected to form in a large fraction of the available parameter space of 2HDMs with a $U(1)_{\text{PQ}}$ symmetry, although only a limited subset of the space has been investigated so far. This variety of superconducting string shares many qualitative features with the prototypical example --- a model with two complex scalar fields and a $U(1) \times U(1)$ symmetry --- and are shown to be stable for a long time in the global theory via numerical simulations.

These solutions would likely have many of the same effects in cosmology as standard superconducting strings, such as the possibility for Vorton formation, which we intend to investigate in a future work. It should be noted, however, that since 2HDMs are an extension of the electroweak theory, these strings and Vortons would be significantly less massive than GUT scale strings, which is the relevant energy scale for a large fraction of topological defects considered in the literature.

It is also worth mentioning that one of the most popular axion models, namely the DFSZ model \cite{DINE1981199,Zhitnitsky:1980tq}, is based upon a two-Higgs-doublet model with an additional scalar field. Superconducting strings in the DFSZ have been considered in \cite{Abe:2020ure}, although they are based upon a different ansatz for the 2HDM fields. The question of whether the string solutions presented in this paper can be straight-forwardly embedded into the DFSZ model is an important one which may have implications for the mass of the axion if the presence of currents travelling along the string network appreciably alters the dynamics \cite{Fukuda:2020kym}.

\section*{Acknowledgements}

We would like to thank Jeff Forshaw for helpful comments.

\bibliography{refs.bib}
\bibliographystyle{unsrt}

\end{document}